\documentclass[12pt]{iopart}

\usepackage{graphicx}
\usepackage{verbatim}
\usepackage{color}
\usepackage{float}

\newcommand{\Heebar}{$^4\overline{\rm He}$}
\newcommand{\Hee}{$^4$He}
\newcommand{\He} {$^3$He}

\newcommand{\pbar}{$\overline{\rm p}$} 
\newcommand{\tbar}{$^3\overline{\rm H}$} 
\newcommand{\dbar}{$\overline{\rm d}$}

\newcommand{\Hebar}{$^3\overline{\rm He}$}

\begin{document}

\title[] {Observation of the antimatter helium-4 nucleus at RHIC}

\author{L.Xue for the STAR Collaboration $^{a,b}$}
\address{$^{a}$ Shanghai Institute of Applied Physics, Chinese Academy of Sciences,P.O. Box 800-204, Shanghai 201800, China}
\address{$^{b}$ Brookhaven National Laboratory, Upton, NY 11973, USA}
\ead{xueliang@sinap.ac.cn}

\begin{abstract}

We present the observation of the \Heebar~ nucleus, the heaviest antinucleus observed to date. In total, 18 \Heebar~ counts were detected at the STAR experiment at RHIC in 10$^{9}$ recorded Au+Au collisions at beam energies of $\sqrt{s_{NN}}$ = 200 GeV and 62 GeV. The background has been estimated, and the misidentification probability is found to be lower than 10$^{-11}$.
\end{abstract}

\submitto{\JPG}

\section{Introduction}
Relativistic heavy ion collisions can produce high temperature and high density matter containing roughly equal numbers of quarks and antiquarks. The environment is uniquely suited for production of light antinuclei, including \pbar, \dbar, \tbar, \Hebar, \Heebar, and anti-hypernuclei.

During the 2010 RHIC runs, with upgraded readout electronics for the Time Projection Chamber (TPC)\cite{TPC} and fully installed barrel Time Of Flight (TOF)\cite{TOF}, STAR is in a good position to study the production of antimatter nuclei. In this paper, we report the measurement of \Heebar~nucleus\cite{AntiAlpha} at STAR. Approximately 10$^{9}$ Au+Au collisions taken  by TPC in 2007 and by TPC and TOF in 2010 are used in this analysis. Preferential selection of events containing tracks with charge $Ze = \pm 2e$ was implemented using a High-Level Trigger (HLT) for data acquired in 2010.  A new variable, $n_{\sigma_{dE/dx}}$, is defined as $n_{\sigma_{dE/dx}}$ = $\frac{1}{R}\ln(\langle dE/dx \rangle/\langle dE/dx \rangle^{B} )$ (where $\langle dE/dx \rangle^{B}$ is the expected value\cite{Bichsel} for a given particle spice, $R$ is the resolution of $\langle dE/dx \rangle$) for particle identification. HLT performs an online fast track reconstruction to select events that have at least one track with $\langle dE/dx \rangle$, $n_{{\sigma_{^3\overline{\rm He}}}}>$ -3. The HLT identifies 70\% of charge-2 tracks by selecting 0.4\% of the total events for express analysis.

\section{\Heebar~Particle identification}
\begin{figure}[H]
\centering
\makebox[0cm]{\includegraphics[width=0.9 \textwidth]{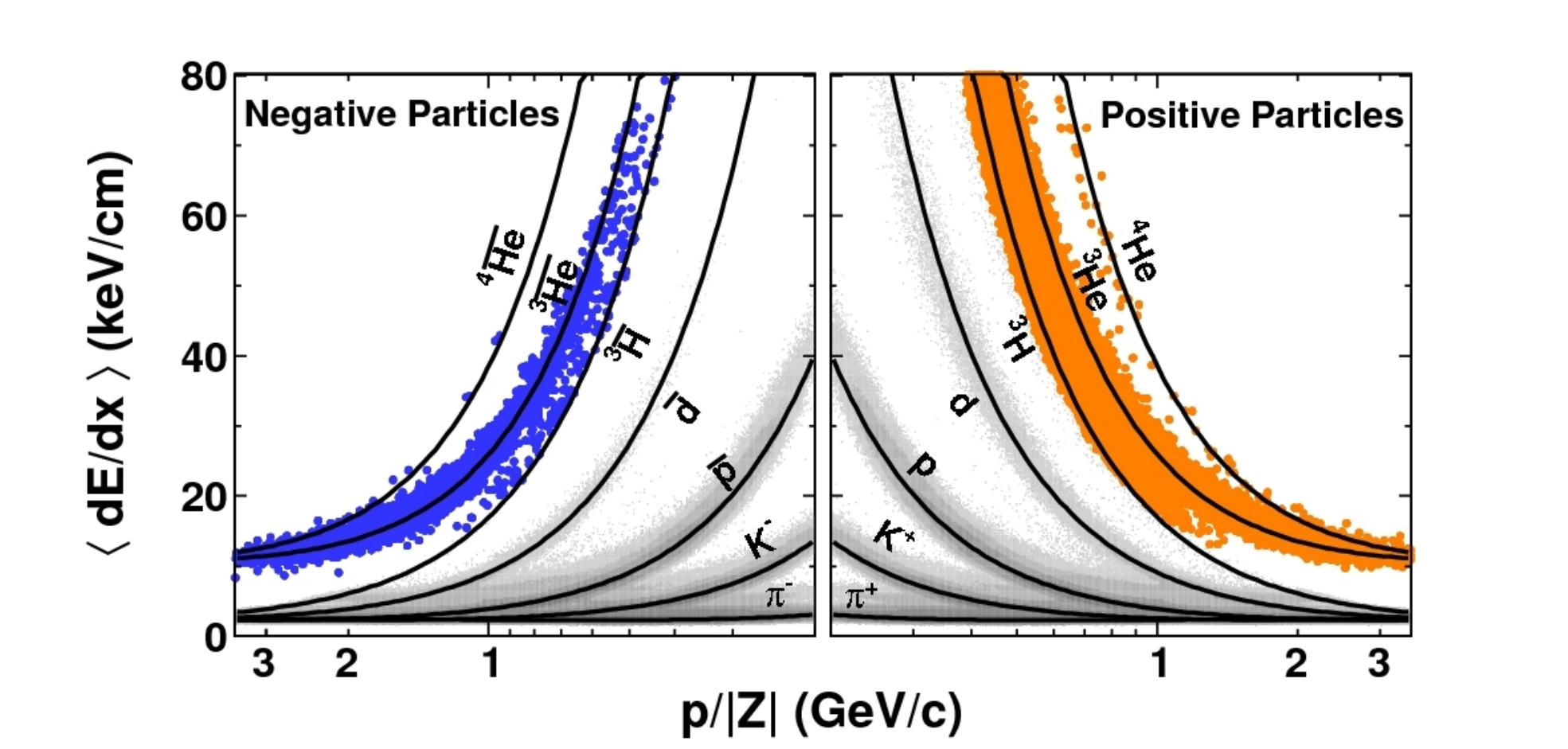}}
\caption{$\langle dE/dx \rangle$ versus $p/|Z|$ for negatively charged particles (left) and positively charged particles (right). The black curves show the expected values for each species. The lower edges of the colored bands correspond to the HLT's online calculation of $3\sigma$ below the $\langle dE/dx \rangle$ band center for \He. For reference, the grey bands indicate the d, \dbar, p, \pbar, K and $\pi$ from Au+Au Minimum bias events at $\sqrt{s_{NN}}$ = 200 GeV.}
\label{fig:dedxvsRigidity}
\end{figure}

TPC is able to reconstruct charged tracks between $\pm$1.8 units of pseudo-rapidity with full azimuthal angle. Tracks between pseudo-rapidity $\pm$1 are used for this analysis. Collisions are required to have primary vertex position within 30 cm from the center of TPC along beam line direction, and have a less than 3 cm difference between z vertex reconstructed by Vertex Position Detector (VPD) and TPC. Particle identification can be achieved by correlating the ionization energy loss $\langle dE/dx\rangle$ of charged particles in TPC gas with their measured magnetic rigidity. A set of selection criteria is employed to get a clean separation between different particle species, including at least 25 ionized electron clusters of tracks in TPC, more than 15 clusters for $\langle dE/dx\rangle$ calculation, and the distance of closest approach (DCA) between tracks and primary vertex less than 3 cm for negative particles ( less than 0.5 cm for positive particles). Figure \ref{fig:dedxvsRigidity} shows $\langle dE/dx \rangle$ versus rigidity ($p/|Z|$) distribution. A distinct band centered around the expected value\cite{Bichsel} for \Hee~ particles is shown in the right panel. In the left panel, where $p/|Z|$ is less than 1.4 GeV/c, 4 \Heebar~ particles are located within the expected band for \Heebar~ and well separated from the \Hebar~ band.

\begin{figure}[H]       
\centering
\makebox[0cm]{\includegraphics[width=0.6 \textwidth]{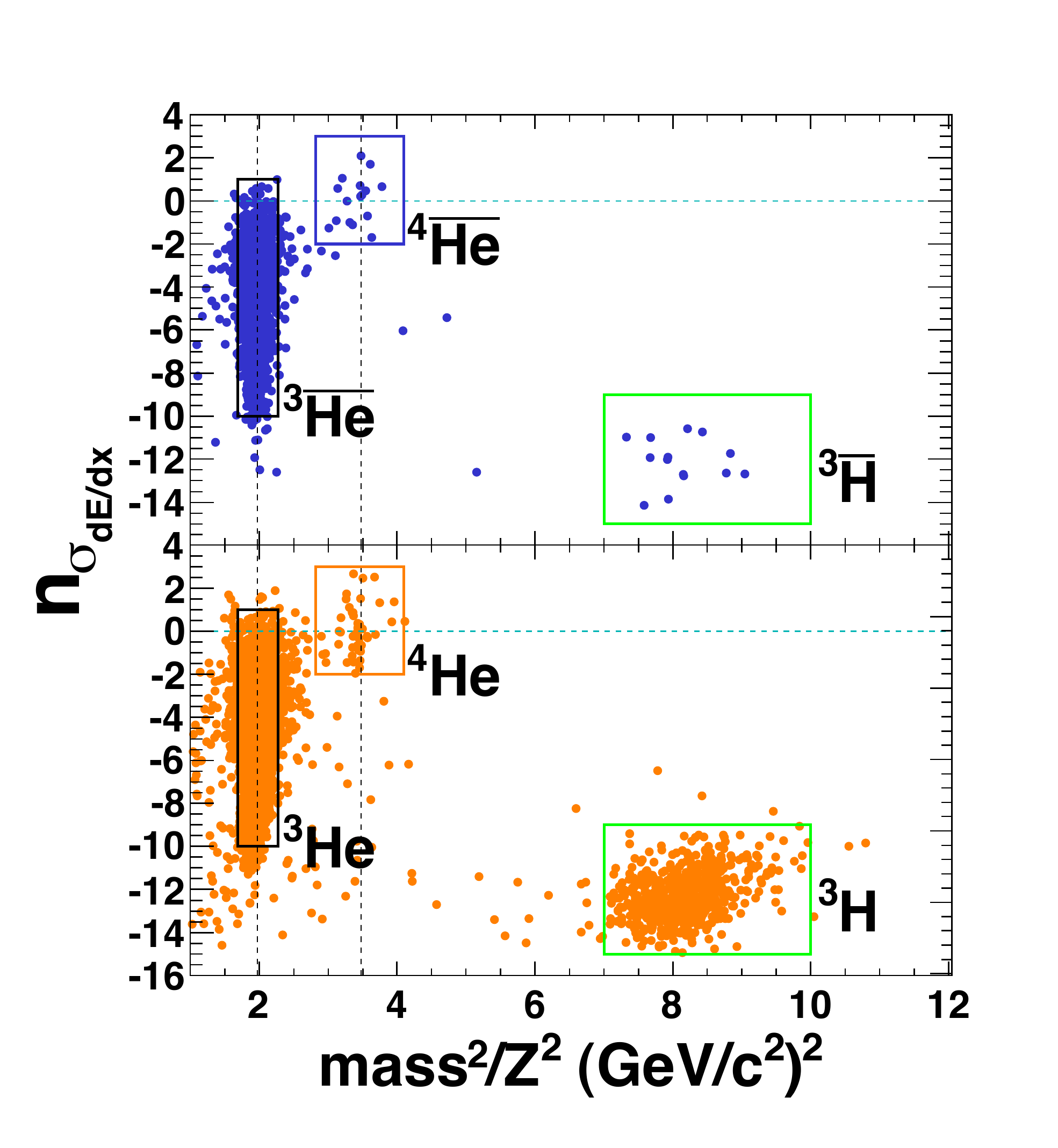}}
\caption{The plot shows the $n_{\sigma_{dE/dx}}$ vs $m^2/Z^2$ distribution of negatively charged particles (top panel) and positively charged particles (bottom panel). The values of $m^2/Z^2$ for \He~(\Hebar) and  \Hee~(\Heebar) are indicated by the vertical lines at 1.97 GeV$^2/c^4$ and 3.47 GeV$^2/c^4$, respectively. The horizontal line marks the position of $n_{\sigma_{dE/dx}}=0$ for \Hee~(\Heebar). }
\label{fig:m2VsNSigma}
\end{figure}

In addition to the mean energy loss per unit track length $\langle dE/dx\rangle$ in the TPC gas and the momentum provided by the track curvature in the magnetic field, the measurement of \Heebar~relies on the time of flight of particles arriving at the TOF surrounding the TPC. In higher momentum region, a combination of $\langle dE/dx \rangle$ and mass from TOF measurement via  $m^2 = p^{2}(t^2/L^2 -1)$ ($t$ and $L$ are the time of flight and path length for \Heebar~, respectively) is used for particle identification. An improved signal to background ratio is furthermore achieved by additional selection criteria on the track extrapolations to the TOF hits. Figure \ref{fig:m2VsNSigma} shows the $n_{\sigma_{dE/dx}}$ versus mass distribution. In the top and bottle panel, the majority species are \Hebar~and \He, and there are \Heebar~and \Hee~particles around $n_{\sigma_{dE/dx}} = 0$ and $m^{2}/Z^{2}$ = 3.47 GeV$^2/c^4$. Other particles in these two panels are $^3$H and \tbar. In total, 18 counts for \Heebar~are observed: 16 have been reconstructed from Au+Au collisions in 2010, and 2 counts have been identified by $\langle dE/dx \rangle$ alone from data recorded in 2007. The latter are not included in this plot because the STAR TOF was not installed at that time.

\section{\Heebar~Background estimation}
\begin{figure}[H]   
\centering
\makebox[0cm]{\includegraphics[width=0.9 \textwidth]{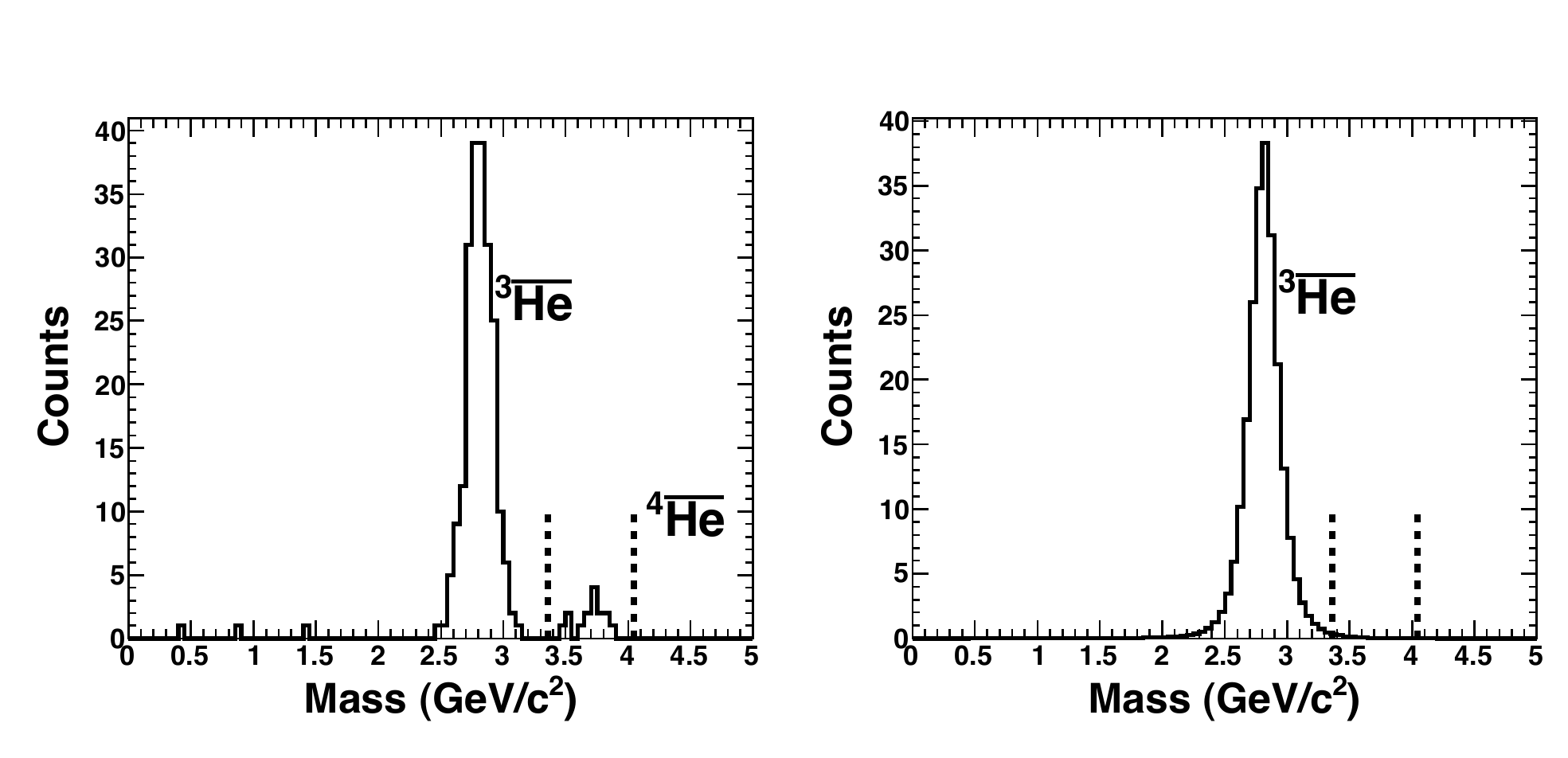}}
\caption{Left panel shows the mass distribution for charge-2 tracks from data, right panel shows the reproduced \Hebar~mass distribution. Dashed lines represent the mass window for \Heebar.}
\label{fig:mass}
\end{figure}

The mass distribution of \Hebar~may extend to \Heebar~selection area because of the finite TOF timing resolution, thus contributing to the background of \Heebar. We reproduce \Hebar~mass distribution with each track's expected time of flight smeared by time deviation ($t-t^{expected}$) of other tracks from the same data sample. Figure \ref{fig:mass} shows the mass distribution for charge-2 tracks from real data (left panel) and reproduced \Hebar~mass distribution (right panel). Then, we obtain the background by integrating reproduced \Hebar~mass distribution over the cut window (3.35 GeV/$c^2 <$ mass $<$ 4.04 GeV/$c^2$) of the \Heebar~ selection. We estimate that the background contributes 1.4 and 0.05 counts of the 15 and 1 counts from the Au+Au collisions at 200 and 62 GeV, respectively. Therefore, the probability of misidentification is calculated in the way of Poisson distribution\cite{Poisson} and is at the $10^{-11}$ level .

\section{Summary}
We present the observation of \Heebar~ nucleus at STAR. In total, 18 \Heebar~counts have been detected. Of those, 2 of them were identified by TPC alone with data recorded 2007, 15 and 1 were identified by both TPC and TOF from 2010 Au+Au collisions at $\sqrt{s_{NN}}$ = 200 GeV and 62 GeV with backgrounds of 1.4 and 0.05. The misidentification probability is estimated to be below 10$^{-11}$.

\section{Acknowledgments}
 This work was supported in part by the NSFC of China under Grant Nos. 11035009, the Shanghai Development Foundation for Science and Technology under Contract No. 09JC1416800, and the Knowledge Innovation Project of the Chinese Academy of Sciences under Grant No. KJCX2-EW-N01.

\end{document}